\documentstyle[epsf]{EuroPhys}
\input EuroMacr
\begin{document}
\euro{}{}{}{}
\Date{12 January 1998}
\shorttitle{L. YIN {\em et. al.\/} SPIN-SPIN CORRELATION LENGTHS OF BILAYER
ANTIFERROMAGNETS}
\title{Spin-spin correlation lengths of bilayer
antiferromagnets}
\author{Lan Yin$^1$, Matthias Troyer$^2$, and Sudip Chakravarty$^1$}
\institute{
           $^1$Department of Physics and Astronomy, University of
California, Los Angeles,
CA 90095, U. S. A.\\$^2$Institute for Solid State Physics, University of
Tokyo, Roppongi 7-22-1, Minatoku, Tokyo 106, Japan}
\rec{}{}
\pacs{
\Pacs{75}{10.Jm}{Quantized spin models}
\Pacs{74}{72}{High T$_c$ cuprates}
\Pacs{74}{72.Bk}{Y-based compounds}
}
\maketitle
\begin{abstract}
The spin-spin correlation length and the static structure factor for
bilayer antiferromagnets, such as YBa$_2$Cu$_3$O$_{6}$, are calculated
using field theoretical and numerical methods. It is shown that these
quantities can be directly measured in neutron scattering experiments
using energy integrated two-axis scan despite the strong intensity
modulation perpendicular to the layers.  Our calculations show that
the correlation length of the bilayer antiferromagnet diverges
considerably more rapidly, as the temperature tends to zero, than the
correlation length of the corresponding single layer antiferromagnet
typified by La$_2$CuO$_4$. This rapid divergence may have important
consequences with respect to magnetic fluctuations of the doped
superconductors.
\end{abstract}
A powerful method to measure the spin-spin correlation length of a
layered magnet is the neutron scattering method known as the energy
integrated two-axis scan (TAS).  In recent years, TAS has been
successfully applied to La$_2$CuO$_4$\cite{Rev}, which is the parent
compound of one of the high temperature superconductors.  This
experimental technique is not readily extendable, however, to a wide
class of high temperature superconductors with close magnetic bilayers
or triple layers within the unit cell; a particularly important
example is ${ \rm YBa_2Cu_3O_{6}}$, which has a close pair of magnetic
planes within the unit cell, but it is otherwise a square lattice spin
$S=1/2$ Heisenberg antiferromagnet.  The reason for the difficulty is
an intensity modulation\cite{Tranquada} with the momentum transfer
perpendicular to the planes. Thus, there have been {\em no} direct
measurements of the correlation length despite considerable discussion
of the importance of antiferromagnetic fluctuations in these
materials.

In the present paper, we use both the field theoretical
approach\cite{CHN} and the numerical quantum Monte Carlo (QMC) loop
algorithm\cite{Loop} to obtain the low temperature properties of
bilayer antiferromagnets. We also show that in the experimentally
relevant regime TAS can be extended to such antiferromagnets.  Thus,
it is hoped that the antiferromagnetic fluctuations can be explored
more thoroughly in future measurements. This is likely to be important
in understanding the magnetic properties of the superconductors
obtained by doping these antferromagnetic parent compounds.

The Heisenberg model for a spin-$S$ bilayer antiferromagnet is
\begin{equation}
H=J_{\parallel}\sum_{\langle ij \rangle,p}{\bf S}_i^{(p)}\cdot{\bf
S}_j^{(p)}+J_{\perp}\sum_i{\bf
S}_i^{(1)}\cdot{\bf S}_i^{(2)}.\label{HAF}
\end{equation}
The sum in the first term is over the nearest neighbor pairs on a square
lattice in each plane,
where the plane index $p$ takes  two values 1 and 2.
The second term represents the coupling between the planes. The exchange
constants $J_{\parallel}$ and
$J_{\perp}$ are both positive.

The low energy, long wavelength properties of the two dimensional
Heisenberg model is well-described by the quantum $O(3)$ nonlinear
$\sigma$-model\cite{CHN}. Here, we shall consider its generalization
to coupled bilayers. The Euclidean action for this system can be
written down on general symmetry grounds\cite{CHN}, but it can also be
derived from a $(1/S)$ expansion\cite{Dotsenko}. The action is
\begin{eqnarray}
 S&=&\int^u_0 dx_0 \int d^2x \Bigg[ \sum_{p=1}^2
	\left({1 \over 2g_0} |\partial_\mu \hat{{ \bf \Omega}}^{(p)}|^2  -
h_0 \sigma^{(p)} \right) \nonumber \\
&+&
  {1 \over 2g_{\perp}^0} |\hat{{ \bf \Omega}}^{(1)}-\hat{{ \bf
\Omega}}^{(2)}|^2 -{1\over2g_t^0}(\hat{{ \bf \Omega}}^{(1)}
	 \times \partial_0 \hat{{ \bf \Omega}}^{(1)}) \cdot (\hat{{ \bf
\Omega}}^{(2)}
	 \times \partial_0 \hat{{ \bf \Omega}}^{(2)})\Bigg] ,
\end{eqnarray}
where $\sigma^{(p)}$ is the component of the staggered field $\hat{{ \bf
\Omega}}^{(p)}$ in the direction of the
staggered magnetic field $h_0$. Here, all the coupling constants and
dimensional variables have been
scaled to their dimensionless forms.  We shall work directly at
$d=2$; the label $\mu =1,2$ denotes the spatial directions, and $\mu=0$
denotes  the imaginary time direction of  extent $u$.

The relations between the $\sigma$-model parameters and the Heisenberg
model parameters
are known in the large-$S$ limit\cite{Dotsenko,BL}, but we shall not need
them in the present papper.  The momentum cutoff of the
$\sigma$-model, $\Lambda$, is chosen to be
$\sqrt{2\pi}/ a$, where $a$ is the lattice constant of the Heisenberg model,
to conserve the number of
degrees of freedom. The bare dimensional bilayer gap is
given by $\Delta_{\perp}^0\equiv\hbar
c_0\Lambda\sqrt{2g_0/g_{\perp}^0}$, where $c_0$ is the bare spin wave
velocity at the scale
$\Lambda^{-1}$.

It can be seen from explicit calculations\cite{IRL} that the angular
momentum coupling between the layers is
irrelevant for weakly coupled bilayer systems. Moreover, this coupling
reduces to a higher
gradient coupling when the
number of layers tends to infinity. Therefore, we shall omit this term  even
though its presence breaks the ``Lorentz invariace" and  slightly renormalizes
the spin-wave velocity.

For notational simplicity, it is
useful to define
$\tilde{h}_0=h_0 g_0$ and $\tilde{\gamma}_{\perp}^0=(2 g_0)/ g_\perp^0$.
The  one-loop
momentum-shell calculations similar to  those of Ref.~\cite{CHN} yields
\begin{eqnarray}
{dg \over dl} &=& -g+{g^2 \over8\pi}\left[F_1(g/2t,\tilde{h})+F_2(g/2t,\tilde{h},\tilde{\gamma}_{\perp})\right], \\
{dt \over dl} &=& {g t \over 8\pi}\left[F_1(g/2t,\tilde{h})+F_2(g/2t,\tilde{h},\tilde{\gamma}_{\perp})\right], \\
\label{dhp}
{d\tilde{\gamma}_{\perp} \over dl} &=& 2\tilde{\gamma}_{\perp}-{g
\tilde{\gamma}_{\perp} \over 4\pi}
F_2(g/2t,\tilde{h},\tilde{\gamma}_{\perp}), \\
{d\tilde{h}\over dl} &=& 2\tilde{h},
\end{eqnarray}
where $e^l$ is the length rescaling factor,
$F_1(g/2t,\tilde{h})=\coth(g\sqrt{1+\tilde{h}}/2t)/\sqrt{1+\tilde{h}}$, and $F_2(g/2t,\tilde{h},\tilde{\gamma}_{\perp})=
\coth(g\sqrt{1+\tilde{h}+\tilde{\gamma}_{\perp}}/2t)/\sqrt{1+\tilde{h
}+\tilde{\gamma}_{\perp}}$.
The variable
$t$ is the dimensionless temperature variable. The dimensionless thickness of the slab in the imaginary
time direction
$u=g/t$ satisfies a simple scaling relation given by
$(g/ t)=(g_0/t_0) e^{-l}$.

\begin{figure}[tb]
\centering
\noindent
\begin{minipage}[t]{.48\linewidth}
\centering
\leavevmode
\epsfxsize = \linewidth
\epsffile{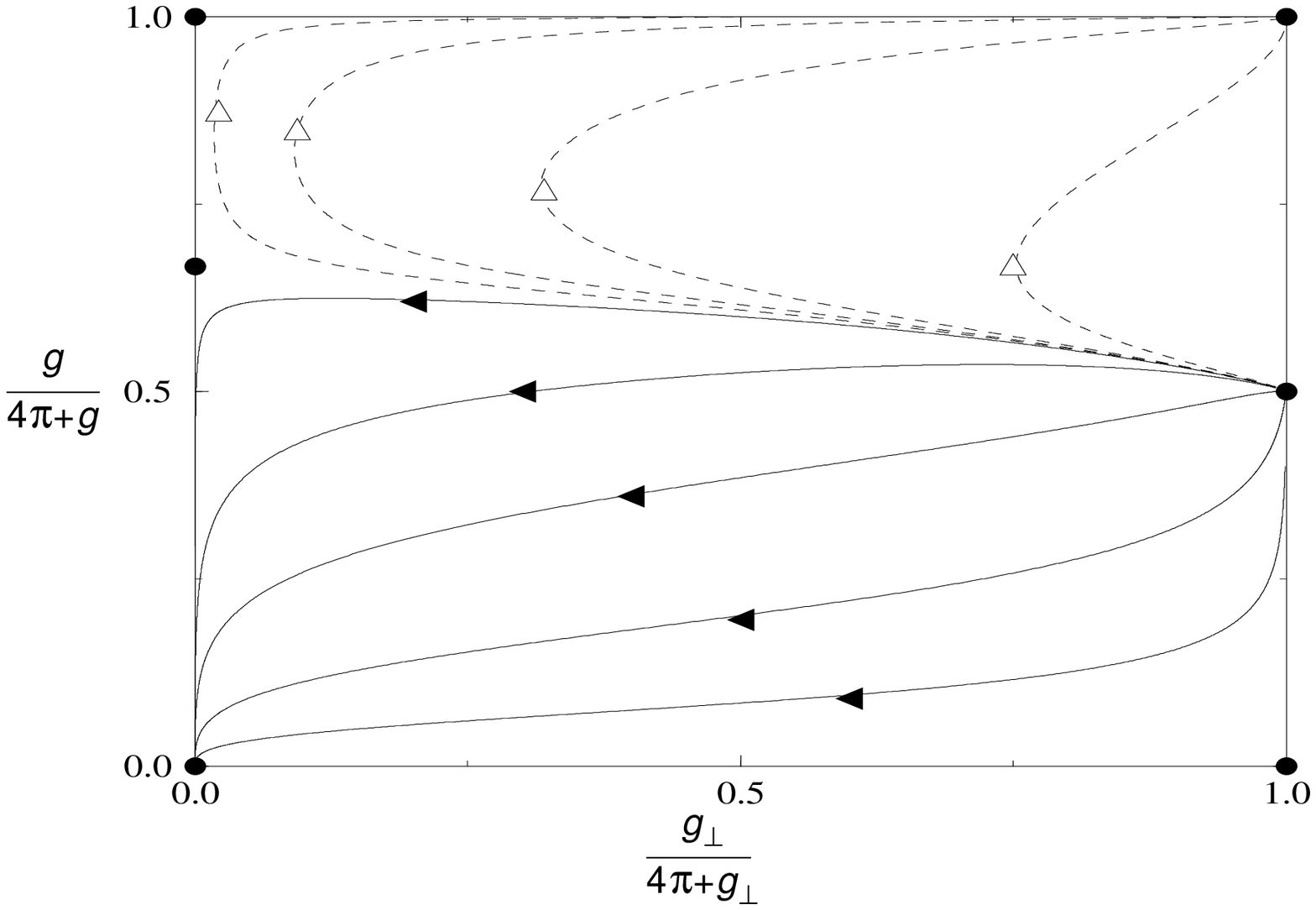}
\label{ZTP}
\end{minipage}
\begin{minipage}[t]{.48\linewidth}
\centering
\leavevmode
\epsfxsize = \linewidth
\epsffile{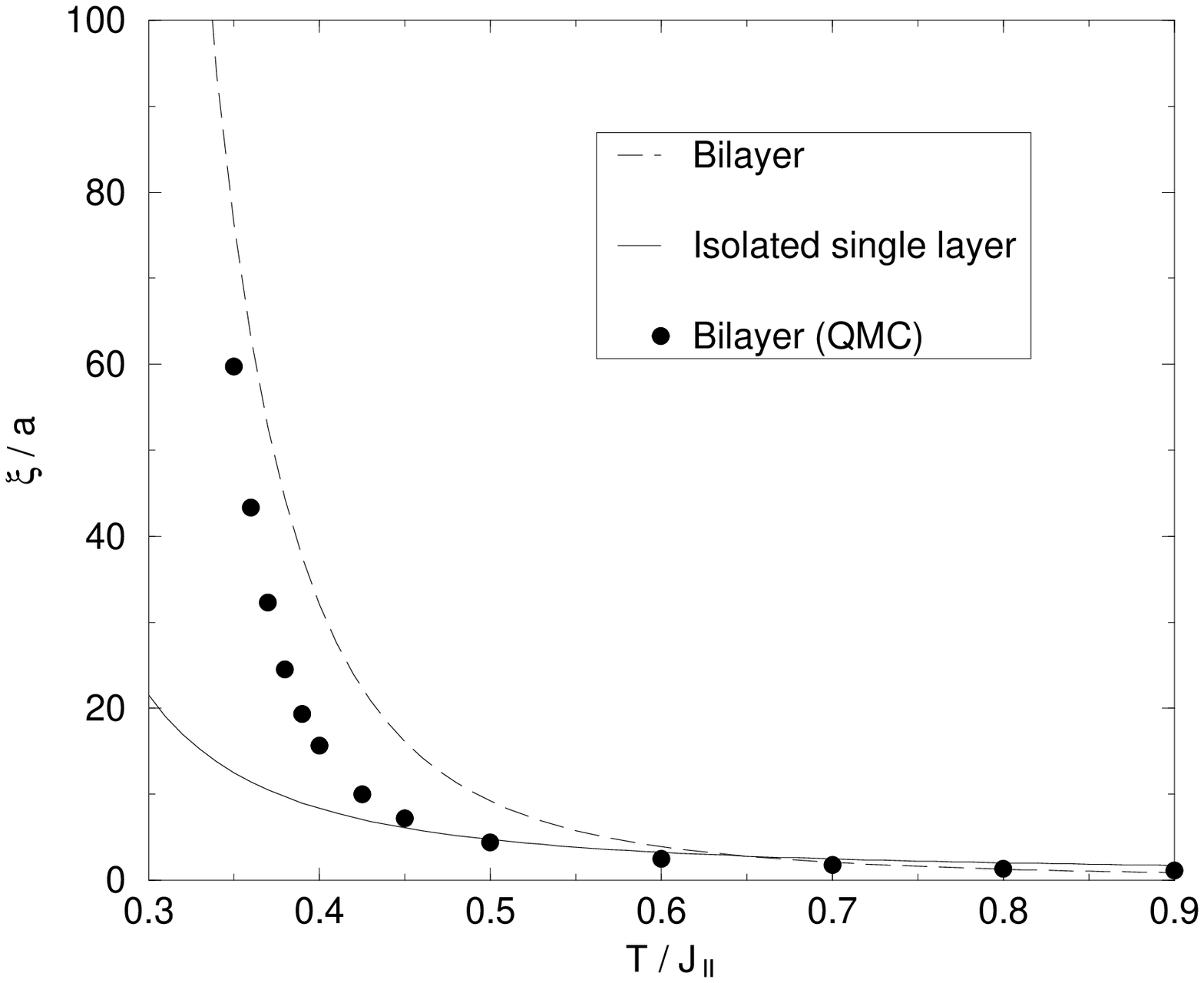}
\label{corr}
\end{minipage}
\caption{Left: Zero temperature flow diagram.}
\caption{Right: Correlation lengths of the bilayer antiferromagnet
compared to those of the isolated
single layer. The solid circles are the Monte Carlo results.}
\end{figure}

The zero temperature flows of the renormalization group equations are shown
in Fig.~1\ref{ZTP}.
There are
two phases, separated by a separatrix between the two unstable fixed points
$(g=4\pi, g_\perp=\infty)$ and
$(g=8\pi,g_\perp=0)$;  the former is the fixed point of the single layer
case\cite{CHN}.  There
are two stable fixed points.  The ordered-phase fixed point is located at
$(0,0)$, where both the
in- and the inter-plane couplings are infinitely strong.  The
disordered-phase fixed point is located at
$(\infty,\infty)$, where the system becomes totally disordered.  Although
quantum nonlinear $\sigma$-model
is a very accurate description of the low energy physics in or near the
ordered phase,
far into the disordered phase, such a continuum theory can not be expected to
be valid.

{}From experiments on ${ \rm YBa_2Cu_3O_{6+x}}$, it is known that the ground
state is an
ordered N{\'e}el state and that
$J_{\perp}\sim 0.1 J_{\parallel}$\cite{OPTM}. Thus, the parameters
are such that  $\tilde{\gamma}_{\perp}^0 \sim (J_\perp /
J_\parallel) \ll 1$, and $g$ is well below the critical value  required for
the phase transition to the
quantum disordered phase at $T=0$. Therefore, the system is in the
renormalized classical
regime\cite{CHN}.

To proceed further, we need an analytical solution of the renormalization
group equations. We have obtained a
good appoximation to the solution  based on the following observations.
In the renormalized classical regime, the bilayer gap
$\sqrt{\tilde{\gamma}_{\perp}}$ is initially much smaller than
unity, but  increases as
$\sqrt{\tilde{\gamma}_{\perp}}
\propto e^{\alpha(l) l/2}$, where $\alpha(l)<2$, but tends to  2 for large
$l$.  We can
therefore consider two regions,
$\tilde{\gamma}_{\perp} \ll 1$ in region (I), and ${d\tilde{\gamma}_{\perp}
\over dl }\simeq 2 \tilde{\gamma}_{\perp}$ in the
region (II). We solve the renormalization group equations separately in
regions (I) and (II), and then join
the two solutions together to get the final answer. The result is
\begin{equation}
{1 \over t_0}-{1 \over t}={1 \over 4\pi}\ln \left[{\sinh^2({g_0 \over
2t_0}) \over \sinh({g_0 \over 2t_0}e^{-l} )
	\sinh\left({g_0 \over
2t_0}\sqrt{e^{-2l}+\tilde{\gamma}_{\perp}^0(1-{g_0 \over
4\pi})}\right)}\right]\label{t}.
\end{equation}
The {\em isolated} single layer result
is trivially recovered when the bilayer gap is much smaller than the
inverse  correlation length, that is,
$\sqrt{\tilde{\gamma}_{\perp}^0} \ll e^{-l}$.
In the limit of large $l$, Eq.~(\ref{t}) has the asymptotic form
$2 / t=2 / t_{ \rm eff} - l / 2\pi$,
where
\begin{equation}
{1\over t_{ \rm eff}}={1\over T}\left[\rho_s^0(1-{g_0\over
4\pi})+{\Delta_{\perp}(0)\over
8\pi}\right]+{1\over 4\pi}\ln\left[{\hbar c\Lambda \over
T}\left(1-e^{-{\Delta_{\perp}(0)/ T}}\right)\right].\label{teff}
\end{equation}
{}From the solution of the renormalization group equations at $T=0$, the
quantity inside the first square
parenthesis on the right hand side of Eq. (\ref{teff}) can be shown to be
the one-loop renormalized
zero temperature bilayer spin stiffness constant. Similarly, one can show
that $\Delta_{\perp}(0)=\hbar
c \Lambda \sqrt{{\tilde \gamma}_{\perp}^0(1-{g_0\over 4\pi})}$ is the
one-loop renormalized bilayer gap
at $T=0$. We assume that these renormalizations hold to all orders. Thus,
we can express $t_{\rm eff}$
in terms of the physical bilayer parameters, denoted by the superscript $b$, as
\begin{equation}
{1\over t_{\rm eff}}={\rho_s^b(0)\over T}+{1\over 4\pi}\ln\left[{\hbar
c\Lambda \over
T}\left(1-e^{-{\Delta_{\perp}^b(0)/ T}}\right)\right]
\end{equation}
The problem is now mapped onto an effective classical $\sigma$-model if we
identify
$2 / t_{ \rm eff}=1 / t^{ \rm cl}_0$ and $2 / t=1 / t^{ \rm cl}$. One can
then show\cite{CHN} that
\begin{equation}
\xi={e \over 8} \Lambda^{-1} {t_{ \rm eff} \over 4\pi} e^{4\pi \over t_{
\rm eff}}\left[1-0.5{t_{ \rm
eff}\over 4\pi}\right]
\label{Xi}
\end{equation}
with the exact prefactor determined in Ref.~\cite{Prefactor}.

The physical parameters can be calculated from the spin wave theory of the
antiferromagnetic
bilayer Heisenberg model in Eq.~(\ref{HAF}). For $S=1/2$,  the spin wave
velocity of the
bilayer complex
calculated  to order $(1/S)$ is given by
$\hbar c=\sqrt{2}J_{\parallel} Z_c a (1+{J_{\perp}\over
4J_{\parallel}})^{1/2}$; similarly,
$\Delta_{\perp}^b(0)= 2\sqrt{J_{\parallel}J_{\perp}}Z_{\Delta}$.
For $J_{\perp}=0.08J_{\parallel}$\cite{OPTM}, $Z_{\Delta}=0.95$ and
$Z_c=1.15$. The bilayer spin
stiffness constant can be calculated from the hydrodynamical
relation\cite{CHN} $\rho_s^b(0)={1\over
4}J_{\parallel}Z_c^2Z_{\chi}$, where the renormalization factor for the
uniform susceptibility
$Z_{\chi}=0.53$ for the same set of parameters.

The static structure  factor contains two pieces, one for the symmetric combination
of the unit vectors from each layer and the other for the antisymmetric
combination.
The notation in the $\sigma$-model is exactly the opposite to the notation
in the Heisenberg model,
since the unit vector field $\hat{{ \bf \Omega}}^{(p)}$ in the
$\sigma$-model is the continuum limit of the
direction vector of the staggered spin operator $(-1)^{i+p} {\bf
S}_i^{(p)}$ in the Heisenberg model.
For the symmetric piece, we find that
$
S_{ \rm s}(k,t_0)=t_0^2 \xi^2 f(x)/ (2 \pi),
$
where $x=k \xi$ and $f(x)=[1+{1 \over 2}\ln(1+x^2)](1+x^2)^{-1}$.
Similarly, the antisymmetric piece is
$
S_{ \rm a}(k,t_0) \simeq t_0 (2-{g_0 \over 2\pi})  (k^2+\xi_\perp^{-2})^{-1}.
$
where $\xi_\perp$ is the length scale associated with the bilayer gap;
approximately, we have $\xi_\perp
=\hbar c/\Delta_{\perp}^b(0)$.
The symmetric piece is clearly dominant in the long-wavelength limit.

In TAS, the wavevector of the incoming neutron ${\bf q}_i$ is fixed, while the
outgoing neutrons in a direction perpendicular to the layers are  collected,
regardless of their energies.  The transferred wavevector is given by ${\bf
q}={\bf q}_f-{\bf q}_i$.  Its in-plane component ${\bf q}_\parallel$ is a
constant,
${\bf q}_\parallel=-{\bf q}_{i\parallel}$, while its perpendicular component is
a variable, $q_\perp=q_f-q_{i\perp}$. For ${\bf q}_\parallel$ near the
reciprocal
lattice vectors, the form factor is approximately a constant, and the intensity
is an integral over the 3D dynamic structure factor, which  is related to
the 2D dynamic structure
factors  by $S^{ 3\rm D}({ \bf q}, \omega)= \sin^2({ q_\perp
h \over 2}) S_{ \rm a}^{ 2\rm D}({\bf q}_{\parallel},\omega)+
\cos^2({ q_\perp h \over 2}) S_{ \rm s}^{ 2\rm D}({\bf q}_{\parallel},
\omega)$,
where $h$ is the distance between the two layers. The quantity $S_{ \rm
a}^{ 2\rm D}({\bf q}_{\parallel},\omega)$ corresponds to  the antisymmetric
spin combination with respect to the layers, {\em symmetric} in the
$\sigma$-model sense,
and $S_{ \rm s}^{ 2\rm D}({\bf q}_{\parallel},\omega)$ to the corresponding
symmetric spin combination, {\em antisymmetric} in the $\sigma$-model
sense.

In experiments one probes the region ${\bf q}_{\parallel}\approx {\bf G}$,
where $\bf G$ is the nearest
antiferromagnetic reciprocal lattice vector. Defining ${\bf k}={\bf
q}_{\parallel}-{\bf G}$, we can
rewrite the intensity in terms of the $\sigma$-model structure factors. The intensity consists of two
parts,
$I({\bf k})=I_{ \rm s}({\bf k})+I_{ \rm a}({\bf k})$,
where
\begin{eqnarray}
I_{ \rm s} ({\bf k}) &\sim& \int_0^\infty dq_f \sin^2\left[{(q_f-q_{i\perp}) h
\over 2}\right]
   S_{ \rm s}^{ 2\rm D{\sigma}}({\bf k}, {\hbar^2 \over 2m}(q_f^2-q_i^2)),\\
I_{ \rm a}({\bf k}) &\sim& \int_0^\infty dq_f \cos^2\left[{(q_f-q_{i\perp}) h
\over 2}\right]
   S_{ \rm a}^{ 2\rm D{\sigma}}({\bf k}, {\hbar^2\over2m}(q_f^2-q_i^2)).
\end{eqnarray}
The $q_\perp$-modulation is unimportant in the critical region.  The reason
is that
$S_{ \rm s}^{ 2\rm D{\sigma}}({\bf k},\omega)$ is dominated by the critical
fluctuations near $\omega=0$,
where both $\sin^2\left({(q_f-q_{i\perp}) h \over 2}\right)$ and
${d\omega\over dq_f}={\hbar^2 \over m} q_f$ are essentially constants.
Therefore, we can pull these factors out of the integal and obtain the
intensity approximately proportional to
the static structure factor, $I_{ \rm s}({\bf k}) \sim {m \over \hbar^2
q_i} \sin^2\left({(q_i-q_{i\perp})\over 2}h\right) \int_{-E_i}^{\infty}
d\omega
S_{ \rm s}^{ 2\rm D{\sigma}}({\bf k},\omega) \sim S_{ \rm s}({\bf k},t_0)$,
where we have reverted to the previous notation  by dropping the
superscripts. The quantity $E_i$ is the incident neutron energy.
For the antisymetric piece, we get an upper bound by neglecting the factor
$\cos^2({ q_\perp h \over 2}$),
which is $\int_0^\infty dq_f S_{ \rm a}^{ 2\rm D{\sigma}}({\bf k}, {\hbar^2
\over 2m}(q_f^2-q_i^2))
\sim {m \over \hbar^2 \langle q_f\rangle } S_{ \rm a}({\bf k},t_0)$,
where $\langle q_f\rangle$ is some average of the wavevector.
Because $S_{ \rm s}(0,t_0) \gg S_{ \rm a}(0,t_0)$, the intensity is
dominated by the contribution
from the symmetric piece. Therefore TAS for a bilayer should yield
information about the symmetric piece, hence the correlation length. The
contribution of the antisymmetric
piece should  result in a small broad background.

The NMR relaxation rate for the in-plane Cu site will be given by\cite{CO}
${1\over T_1}\propto T^{3\over 2}\xi$, because  the symmetric structure factor
will dominate when the correlation length $\xi$ is large. Thus, the formula
for the
correlation length can also be tested in NMR experiments, as for
the single layer La$_2$CuO$_4$\cite{Imai}.

For single layer antiferromagnets, QMC simulations have shown that the
field theoretical expression for the correlation length is accurate
for only very large correlation lengths, often larger than those
accessible in experiments\cite{Kim1}. To obtain results that are also
reliable at intermediate and high temperatures we use QMC\cite{Loop}
to calculate the correlation length defined from the second moment of
the structure factor.  As a function of the system size, the results
converge within our statistical errors for systems of linear dimension
$L$ larger than $7\xi$. We take this into account in estimating the
infinite volume correlation length. 

The definition of the correlation length from the second moment of the
static structure factor is not equivalent to that used in the field
theoretical approach. The reason is the logarithmic correction $1+{1
\over 2}B_f\ln(1+x^2)$ to the Lorentzian form, where $B_f=1$ in one
loop order. However, from previous simulations, it is known
that $B_f$ is strongly renormalized  and that the
actual correction is an order of magnitude smaller\cite{Tyc}. The two
definitions of the correlation length thus differ by only a few
percent.

In Table 1 we present our results for an inter layer coupling of
$J_{\perp}=0.08 J_{\parallel}$
and plot them in Fig. 2 together with the field theoretical results.
\begin{table}[h]
\caption{Second moment correlation length of the bilayer antiferromagnet at an
inter layer coupling $J_{\perp}=0.08 J_{\parallel}$ calculated in quantum Monte
Carlo simulations.}
\begin{tabular}{cccccccccc}
\hline\\
$T/J_{\parallel}$ & 0.35&0.36 & 0.38 & 0.4&0.45&0.5&0.6&0.7&1\\
$\xi$&	59.7(4)&
43.3(3)&
24.5(2)&
15.6(1)&
 7.14(7)&
 4.37(4)&
 2.46(2)&
 1.73(1)&
0.96(1)\\
\hline
\end{tabular}
\end{table}
Not unexpectedly there are differences at intermediate temperatures,
but the correlation length rapidly approaches Eq. (\ref{Xi}) at lower
temperatures. We find it surprising that at temperatures as high as
$5J_{\perp} (=0.4 J_{\parallel})$ the correlation length of the
bilayer antiferromagnet is already significantly higher than the
correlation length of the single layer antiferromagnet\cite{Sandvik}.

In summary, the theoretical results obtained here should be important
in understanding the magnetic fluctuations of bilayer high temperature
superconductors, as they can be explored in neutron scattering
experiments using the energy integrated two axis scan.   The
applicability of this technique is due to the fact that the spin-spin
correlation length at low temperatures is exceedingly long in bilayer
antiferromagnets\cite{Yin}, and nearly critical fluctuations dominate
for the parameter regime that is of experimental interest. The energy
integration is thus correctly carried out. The present work shows that the bilayer complex in high temperatures such as YBCO are effectively strongly coupled despite the fact that $J_{\perp}\sim 0.1 J_{\parallel}$.

\stars

We thank O. Sylju{\aa}sen for interesting discussions.  This work was
supported by the National Science Foundation, Grant. No. DMR-9531575.
The QMC calculations were performed on the Hitachi SR2201 massively
parallel computer of the computer center at the University of Tokyo.


\begin{thebibliography}{99}
\bibitem{IRL} Defining $\tilde{\delta}_t=g_0/(2g_t^0)$, the renormalization
group equation for the
angular momentum coupling is given by
     $d\tilde{\delta}_t/dl=-(\tilde{\delta}_tg/4\pi) \left[ \coth \left({g
\over 2t}
     \sqrt{{1+\tilde{h}+\tilde{\gamma}_{\perp} \over 1+\tilde{\delta}_t}}
\right) /
     \sqrt{(1+\tilde{h}+\tilde{\gamma}_{\perp})(1+\tilde{\delta}_t)} \right].$
     Since $\tilde{\delta}_t$ always decreases and its initial value is
very small,
     $\tilde{\delta}_t^0 \propto {J_\perp \over J_\parallel} \ll1$, it is
irrelevant.
\end{thebibliography}
\end{document}